\documentclass[aps,prb,twocolumn,superscriptaddress,showpacs,floatfix]{revtex4}

\usepackage{epsfig}

\begin{document}

\title  {Superconducting strip in an oblique magnetic field}

\author{G.~P.~Mikitik}
\affiliation{Max-Planck-Institut f\"ur Metallforschung,
   D-70506 Stuttgart, Germany}
\affiliation{B.~Verkin Institute for Low Temperature Physics
   \& Engineering, Ukrainian Academy of Sciences,
   Kharkov 61103, Ukraine}

\author{E.~H.~Brandt}
\affiliation{Max-Planck-Institut f\"ur Metallforschung,
   D-70506 Stuttgart, Germany}

\author{M.~Indenbom}
\affiliation{Institute of Solid State Physics, Chernogolovka,
Russia}

\date{\today}

\begin{abstract}
As an example for a seemingly simple but actually intricate problem,
we study the Bean critical state in a superconducting strip of
finite thickness $d$ and width $2w \gg d$ placed in an oblique
magnetic field. The analytical solution is obtained to leading
order in the small parameter $d/w$. The critical state depends on
how the applied magnetic field is switched on, e.g., at constant
tilt angle, or first the perpendicular and then the parallel field
component. For these two basic scenarios we obtain the distributions
of current density and magnetic field in the critical states. In
particular, we find the shapes of the flux-free core and of the
lines separating regions with opposite direction of the critical
currents, the detailed magnetic field lines (along the vortex lines),
and both components of the magnetic moment. The component of the
magnetic moment parallel to the strip plane is a nonmonotonic
function of the applied magnetic field.
\end{abstract}


\maketitle

\section{Introduction}

Platelet-like type-II superconductors in a magnetic field applied
at some angle $\theta$ to the normal of their plane are frequently
investigated in various experiments, see, e.g.,
Refs.~\onlinecite{i1,i2,z,t,a}. However, even for the simplest
case of an infinitely long strip placed in an oblique magnetic
field, the critical state was theoretically studied only in the
situation when the magnitude of the applied magnetic field $H_a$
considerably exceeds the field of full-flux penetration into the
sample,\cite{z} $H_p$. The attempt to investigate the critical
state in fields $H_a\le H_p$ led to incorrect results, \cite{b}
since an essential feature of this state was
overlooked, as will be evident from our analysis in Sec.~II B.

In this paper we consider the following basic situation: A thin
superconducting strip fills the space $|x| \le w$, $|y| < \infty$,
$|z| \le d/2$ with $d\ll w$; a constant and homogeneous external
magnetic field $H_a$ is applied at an angle $\theta$ to the $z$
axis ($H_{ax}=H_a\sin\theta$, $H_{ay}=0$, $H_{az}=H_a\cos\theta$).
It is assumed that the thickness of the strip, $d$, exceeds the
London penetration depth, the critical current density $j_c$ does
not depend on the local induction $B$ (Bean model\cite{13,14}),
and the lower
critical field $H_{c1}$ is sufficiently small so that we may take
$B= \mu_0 H$. We consider two scenarios of switching on the
external magnetic field: First, the magnitude of the external
field increases from $0$ to $H_a$ at a fixed angle $\theta$;
second, one turns on $H_{az}$ first and then $H_{ax}$.
Interestingly, these scenarios lead to different critical states.

Taking into account the result of Ref.~\onlinecite{1} (see also
Refs.~\onlinecite{2,lon}), the smallness of the parameter $d/w$
enables us to split the two-dimensional critical state problem for
the strip of finite thickness into two simpler problems: A
one-dimensional problem across the thickness of the sample, and a
problem for the infinitely thin strip. This splitting becomes
possible since under the condition $d/w\ll 1$ the magnetic fields
and currents in the critical state essentially change along the $x$
direction only on scales which considerably exceed the thickness $d$.

The solution of the critical state problem for the infinitely thin
strip is known.\cite{2a,3,4} The $z$ component of the magnetic field
is completely screened by the currents flowing in the region
$|x|<a$, i.e., one has $H_z=0$ there. The length $a$ is described
by the simple formula:
 \begin{equation} \label{1}
 a(h)={1\over \cosh(h\cos\theta)}.
 \end{equation}
Here and below $h\equiv H_a/H_c$, $H_c=J_c/\pi$, $J_c=j_c d$, and
$w$ is taken as the unit of length ($w=1$). In this region of the
strip, $|x|<a$, the sheet current
$J(x)=\int_{-d/2}^{d/2} j(x,z) dz$ (with current density $j$
along $y$) is given by
\begin{equation} \label{2}
 J(x)=-{2\over\pi}J_c\, {\rm arctan}
    {x \sqrt{1-a^2} \over \sqrt{a^2-x^2}} \,.
\end{equation}
On the other hand, at $a< |x| <1$ where $H_z(x)\neq 0$, one has
$J=-{\rm sign}(x)J_c$ with ${\rm sign}(x)=\pm 1$ for $x>0$ and
$x<0$, respectively. The explicit form of $H_z(x)$ in this region
of the strip is presented in Appendix A.

In the region $|x|<a$ of the real strip (with $d\neq 0$), the flux
lines are practically parallel to the planes of the strip and
penetrate into the sample across its thickness from the upper and
lower surfaces of the superconductor. The penetrating flux fronts
form the boundary of a flux-free core, $z_{\gamma}(x)$, which
thus consists of an upper and a lower branch. Below we consider
only the upper branch since from symmetry considerations one has:
$-z^{\rm lower}_{\gamma}(-x)=z^{\rm upper}_{\gamma}(x)\equiv
z_{\gamma}(x)$.

Following our idea of splitting the critical state problem, we
consider a small section of the strip around an arbitrary point
$x$ ($|x|<a$) as an ``infinite'' slab of thickness $d$ placed in a
parallel dc magnetic field $H_{ax}$ and carrying a sheet current
$J(x)$, Eq.~(\ref{2}). The critical state in such a slab is well
known,\cite{3,4} and this enables us to find the flux fronts, the
distribution of the magnetic fields, and the currents across the
thickness of the strip in the region $|x|<a$. Since the critical
state in the slab depends on how $H_{ax}$ and $J$ was turned on,
the above-mentioned dependence of the critical state in the strip
on the prehistory of $H_{ax}$ and of $H_{az}$ appears. Of course,
a similar procedure may be used in the region $a<|x|<1$ to find
the distribution of the magnetic fields, but there the appropriate
analysis is trivial since $j(x,z)$ is constant; we thus do not
discuss it below.

\section{Magnetic Field is Increased at Constant Tilt Angle} 

In the case of the first scenario of switching on the magnetic
field when $\theta=$const and $h$ has increased monotonically, it
is convenient to introduce the function,
\begin{equation} \label{3}
 F(x,h)\equiv {2\over \pi} \left[ h\sin\theta - {\rm arctan}
  {x \sqrt{1-a(h)^2} \over \sqrt{a(h)^2-x^2}} \right] ,
\end{equation}
which is proportional to the $x$ component of the magnetic field
on the upper surface of the strip at $|x|\le a$, $H_{\rm
us}(x)=H_{ax}+0.5J(x)=(J_c/2)F(x,h)$. Below we shall also use the
two characteristic fields:
\begin{eqnarray} \label{4}
 h_p={\pi \over 2\sin\theta}\,,
\end{eqnarray}
and $h_f$ defined by the equations:
\begin{eqnarray} \label{5}
  h_f=h_p - u +{{\rm arctan}[\tan\theta \,
 {\rm th}(u\cos\theta)] \over \sin\theta}, \nonumber \\
 \cosh(u\cos\theta)=\sin\theta \cosh(h_f\cos\theta),
\end{eqnarray}
where $u$ is some parameter. The meaning of these fields will
become clear below.

\subsection{Interval $0<h<h_f$}   

Consider first the flux front in the interval $0<h<h_f$. It is
essential that there exists a point on the upper plane of the
strip where the derivative $dH_{\rm us}(x)/dx$ vanishes. A simple
calculation gives that this occurs at the point with the
coordinate $x_1$,
\begin{equation} \label{6}
x_1(h)=a(h) \sin\theta.
\end{equation}
Thus, when $h$ increases, the flux lines at $-a(h)<x<x_1(h)$
monotonically penetrate into the strip through its upper surface,
and the shape of the core in this interval of $x$ is determined by
the equation:
\begin{equation} \label{7}
 z_{\gamma}(x)=(d/2)[1-F(x,h)].
\end{equation}
This formula follows from the well-known distribution of the
magnetic field in the ``slab'' shown in Fig.~1 (``Bean
profiles''). On the other hand, in the interval $x_1(h)<x<x_0(h)$
the flux lines {\it leave} the sample, and at $x>x_0$ vortices of
opposite sign penetrate into the strip, Fig.~1. Here the point
$x_0(h)$ is found from the condition $H_{\rm us}(x_0)=0$, i.e.,
from
\begin{equation}\label{8}
F(x_0,h)=0.
\end{equation}
It is clear from the inspection of Fig.~1 that in the interval
$x_1(h) <x <x_2(h)$ the shape of the core is determined by the
flux front occurring at the field $h_*$ which has to be found from
the equation
\begin{equation} \label{9}
 x=a(h_*) \sin\theta.
\end{equation}
Thus, in this interval one has
\begin{equation} \label{10}
 z_{\gamma}(x)=(d/2)[1-F(x,h_*)].
\end{equation}
In the same interval there is also a front $z_1(x)$ separating the
regions of the strip with opposite signs of the critical
current density, see Figs.~1, \ref{fig2}. This front is described
by the formula:
\begin{eqnarray} \label{11}
 z_1(x)=(d/4)[2+ F(x,h) -F(x,h_*) ] \,.
\end{eqnarray}
At the point $x_2(h)$ the front $z_1(x)$ reaches the boundary of
the core $z_{\gamma}(x)$, and hence, this $x_2(h)$ is determined
by the condition $z_{\gamma}(x_2)=z_1(x_2)$. Using formulas
(\ref{9})-(\ref{11}), one then obtains for $x_2$:
\begin{eqnarray} \label{12}
 x_2=a(u) \sin\theta , \nonumber \\
 F(x_2,h)+F(x_2,u)=0.
\end{eqnarray}
At $x_2(h) < x <a(h)$ the critical current density has only
negative sign, $j=-j_c$, at $z>z_{\gamma}$, and we arrive at
\begin{equation} \label{13}
 z_{\gamma}(x)=(d/2)[1+F(x,h)].
\end{equation}

We see that in the interval of the magnetic fields $0<h<h_f$ the
width of the flux-free core is equal to $2a(h)$, but its size
along z is less than $d$, see Fig.~2. When $h \to h_f$, the
difference $a(h)-x_2(h)$ tends to zero, and at $h = h_f$ one has
$x_2(h)=a(h)$. With the use of Eqs.~(\ref{1}) and (\ref{12}) this
condition may be rewritten as Eqs.~(\ref{5}).

\subsection{Interval $h_f <h <h_p$}   

When $h_f <h <h_p$, the upper and the lower branches of the
flux-free core merge in the intervals $x_3<|x|<a$, and hence the
size of the flux-free core in the $x$ direction, $2x_3(h)$,
becomes less than $2a(h)$, see Figs.~3, 4. Here $x_3(h)$, which
lies between $x_1(h)$ and $a(h)$, is determined by the condition,
$z_{\gamma}^{\rm upper}(x_3)=z_{\gamma}^{\rm lower}(x_3)$, or in
the explicit form by the equations:
\begin{eqnarray} \label{14}
 x_3=a(u) \sin\theta , \nonumber \\
 F(-x_3,h)+F(x_3,u)=2,
\end{eqnarray}
where we have taken into account the symmetry of the flux-free
core and formulas (\ref{7}), (\ref{9}), (\ref{10}). In other
words, we find that not only the $z$-size of the core is less than
$d$, but also its $x$-size is less than the width of the region in
the strip, $2a(h)$, where $H_z=0$. In the interval
$-x_3(h)<x<x_1(h)$, the core is again described by Eq.~(\ref{7}),
while in the interval $x_1(h)<x<x_3(h)$ it is given by
Eqs.~(\ref{9}), (\ref{10}). When $h$ becomes equal to $h_p$
defined by formula (\ref{4}), the point $x_3$ reaches $x_1$, and
moreover, the core disappears at all since at this field the
difference $z_{\gamma}^{\rm upper}(x)-z_{\gamma}^{\rm lower}(x)$
vanishes even for $|x|<x_1$. Thus, $h_p=0.5\pi/\sin\theta$ is the
field of full penetration of flux into the strip in the oblique
magnetic field. This field has a simple meaning. In usual
units one finds for the $x$ component of the penetration field,
$H_c h_p\sin\theta = J_c/2$, i.e., the penetration
occurs when the $x$ component of the applied field has completely
penetrated into the slab.

 \begin{figure}  
\epsfxsize= .7\hsize  \vskip 1.0\baselineskip \centerline{
\epsffile{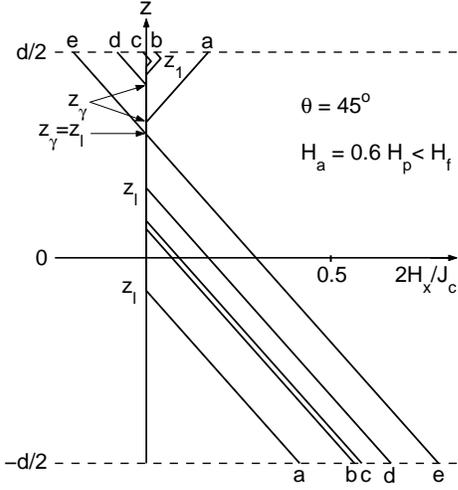}}  \vspace{.1cm}
\caption{\label{fig1}  Bean profiles of the magnetic field
$H_x(z)$ across the strip at 5 positions $x$ in the strip at $h <
h_f$. Here $\theta=45^\circ$, $H_a =0.6 H_p < H_f =0.834 H_p$,
thus $x_1=0.478$, $x_0=0.597$, $x_2=0.617$, $a=0.677$ in units $w$
(see Fig.~6). The characteristic $z$ values $z_\gamma(x)$,
$z_l(x)=-z_\gamma(-x)$, and $z_1(x)$ are defined in the text and
in Fig.~2. Shown are: a) $ 0   < x=0.35 < x_1$, b) $ x_1 < x=0.57
< x_0$, c) $ x_0 < x=0.61 < x_2$, d) $ x_2 < x=0.65 < a  $,  e) $x
\ge a=0.677$. For all these profiles, $2H_x/J_c$ is $F(x,h)$ on
the upper surface, and $F(-x,h)$ on the lower surface. The
continuation of the increasing parts of the profiles b) and c)
intersects the upper surface at $F(x,h_*)$.
 } \end{figure}   

 \begin{figure}  
\epsfxsize= .98\hsize  \vskip 1.0\baselineskip \centerline{
\epsffile{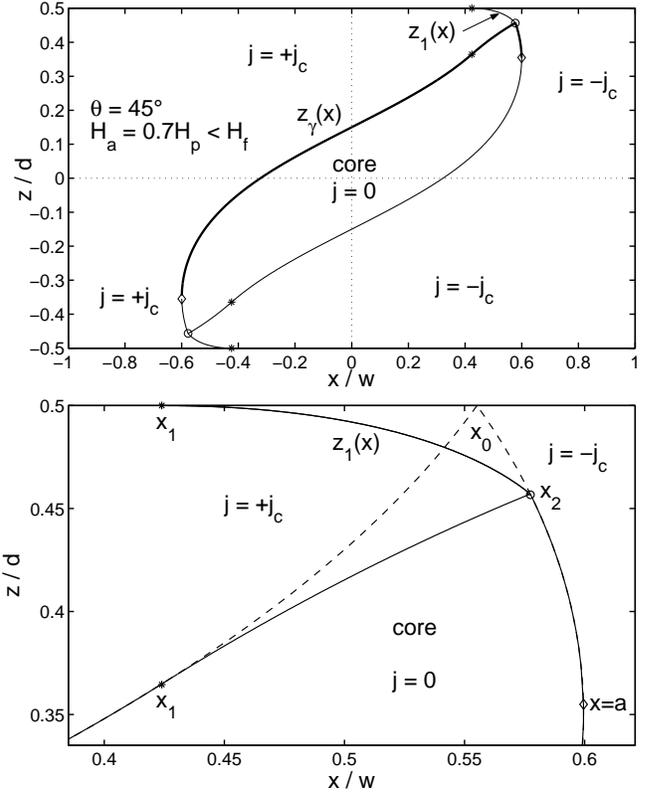}}  \vspace{.1cm}
\caption{\label{fig2} The flux and current fronts in a thin strip
in an increasing magnetic field $H_a$ that is inclined by a
constant angle $\theta=45^\circ$ away from the normal of the strip
plane. Shown is the state when $H_a = 0.7H_p < H_f =0.834 H_p$
where $H_p = (J_c/2)/\sin\theta$ is the field of full penetration,
see text. The current-free core is delimited by the functions
$z_\gamma(x)$ (bold line) and $-z_\gamma(-x)$. The current density
to the left of this core and of the ``tails" $z_1(x)$, $-z_1(-x)$,
is $j=j_c$, and to the right $j=-j_c$. The lower plot enlarges the
region near the points $x=x_1$, $x=x_2$, and $x=a$ on these
curves, see text. The two dashed lines are extensions of the
fronts depicted in the regions $-a\le x \le x_1$ and $x_2\le x \le
a$ and cut the upper surface $z=d/2$ at the point $x=x_0$ where
the local $H_x$ vanishes, Eq.~(8).
 } \end{figure}   

 \begin{figure}  
\epsfxsize= .7\hsize  \vskip 1.0\baselineskip \centerline{
\epsffile{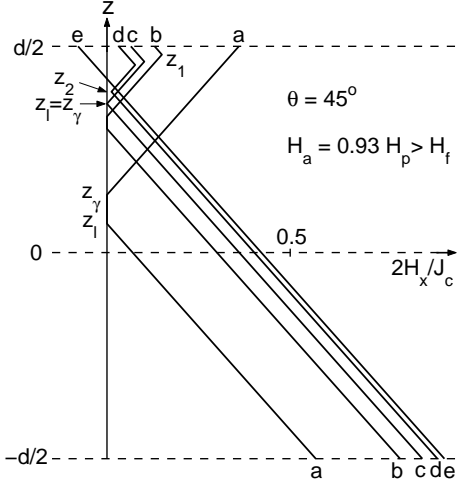}}  \vspace{.1cm}
\caption{\label{fig3} Bean profiles of the magnetic field $H_x(z)$
across the same strip as in Fig.~1 but for higher applied field $h
> h_f$, $H_a =0.93 H_p > H_f =0.834 H_p$, thus
$x_1=0.311$, $x_3=0.422$, $x_0=0.438$, $a=0.440$ in units $w$ (see
Fig.~4). The characteristic $z$ values $z_\gamma(x)$,
$z_l(x)=-z_\gamma(-x)$, $z_1(x)$, and $z_2(x)$ are defined in the
text and in Fig.~4. Shown are the 5 profiles: a) $0 < x=0.15 <
x_1$, b) $x_1 < x=0.40 < x_3$, c) $ x =x_3 =0.422$, d) $x_3 <
x=0.43 < a$,  e) $x \ge a=0.440$. Profile d) does not cut the
field-free core and has $H_x(z)>0$ everywhere. The values of $H_x$
on the upper and lower surfaces are given by the same expressions
$F(x,h)$ and $F(-x,h)$  as in Fig.~1.
 } \end{figure}   

 \begin{figure}  
\epsfxsize= .98\hsize  \vskip 1.0\baselineskip \centerline{
\epsffile{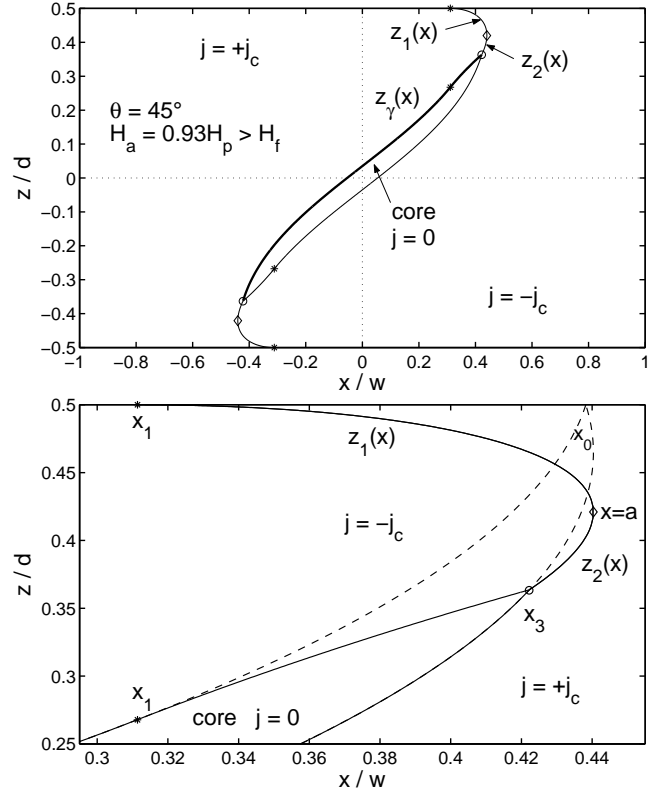}}  \vspace{.1cm}
\caption{\label{fig4} The fronts as in Fig.~2 but at a larger
field $H_a = 0.93 H_p > H_f$. As in Fig.~2 the flux-free core
is formed by the curves $z_\gamma(x)$, $-z_\gamma(-x)$, but the
tail separating regions with $j= \pm j_c$ is now composed of two
functions $z_1(x)$ and $z_2(x)$, which join vertically at
$x=a$ and reach the core at $x=x_3$.
 } \end{figure}   

Interestingly, the part of the boundary of the core in the
interval $x_1(h)<x<x_3(h)$, as well as its part in the interval
$x_1(h) <x <x_2(h)$ for the fields $0<h<h_f$, is described by a
universal function of $z$ on $x$ which does not depend on $h$ at
all, see Eqs.~(\ref{9}), (\ref{10}). The upper corner of the core,
i.e., the point $x_2(h)$ for $0<h<h_f$ or the point $x_3(h)$ for
$h_f <h <h_p$, moves just along the line described by this
function when $h$ increases, Fig.~5.

As to the front separating the regions of the strip with opposite
signs of the critical current density, it is still described by
Eq.~(\ref{11}) in the region $x_1<x<a$ even for $h_f <h <h_p$.
However, in the interval $x_3 \le x \le a$, apart from this upper
branch of the front, $z_1(x)$, a lower branch $z_2(x)$ appears,
and these branches join each other with vertical slope at
$x=a(h)$, Fig.~4. Knowing the upper branch, one can find the lower
branch from the given value of the sheet current $J(x)$,
Eq.~(\ref{2}), yielding
\begin{eqnarray} \label{15}
 z_2(x)=(d/4)[F(-x,h) -F(x,h_*) ] \,,
\end{eqnarray}
where $h_*(x)$ is found from
\[
x=a(h_*) \sin\theta.
\]

A complete set of flux and current fronts is shown in
Fig.~\ref{fig6} for three tilt angles $\theta=30^\circ$,
$45^\circ$, and $60^\circ$. Note that when $H_a$ increases, not
only does the flux-free core shrink but also the current front
separating the regions with $j=\pm j_c$ shifts in the sample. In
other words, the current distribution changes not only near the
core but also in the region away from it. This result disproves
the main assumption of Ref.~\onlinecite{b} that the currents can
change only at the flux front but the critical currents remain
unchanged in the regions penetrated by flux lines.
In Ref.~\onlinecite{b} the flux fronts in inclined field are
thus incorrect.

 \begin{figure}  
\epsfxsize= .98\hsize  \vskip 1.0\baselineskip \centerline{
\epsffile{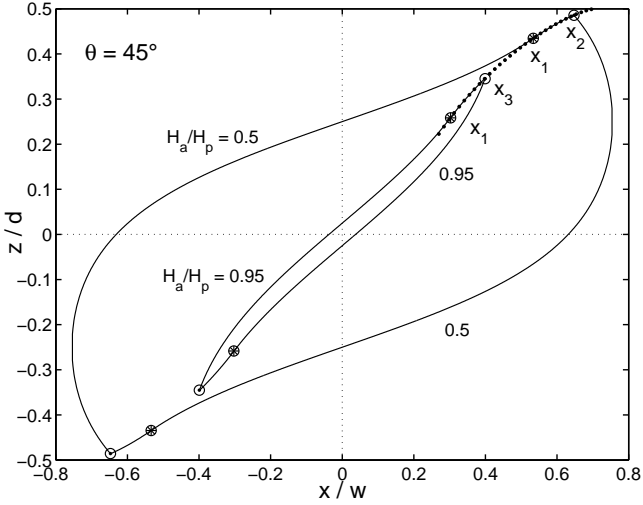}}  \vspace{.1cm}
\caption{\label{fig5} Two flux-free cores in a
thin strip in increasing magnetic field $H_a$ inclined by
$\theta=45^\circ$ as in Figs.~2, 4. The bold dotted line
shows the universal curve, Eqs.~(9,10), on which the upper part
$x_1 \le x \le x_2$ or $x_3$ of all flux fronts lies when
$H_a \le H_p$, see text. The depicted cores belong to
$H_a/H_p = 0.5$ and $H_a/H_p =0.95$. The circles mark the
characteristic points $x_1$, $x_2$, and $x_3$ on the fronts.
 } \end{figure}   

 \begin{figure}  
\epsfxsize= .98\hsize  \vskip 1.0\baselineskip \centerline{
\epsffile{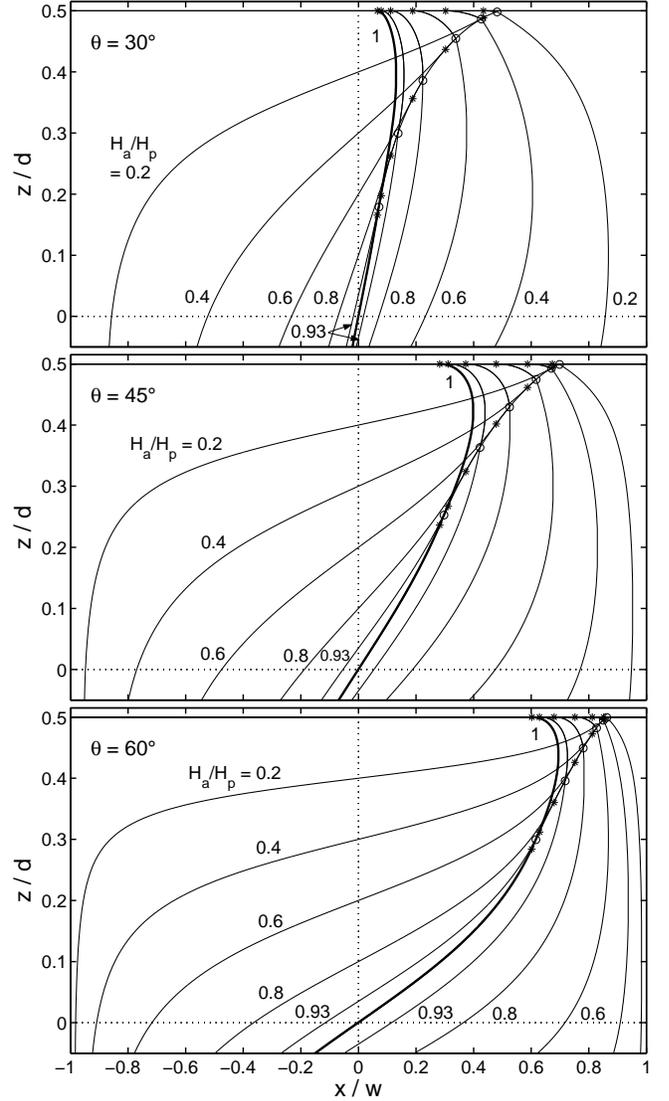}}  \vspace{.1cm}   
\caption{\label{fig6} Flux and current fronts in a thin strip in
an increasing magnetic field $H_a$ inclined by three angles
$\theta=30^\circ$, $\theta=45^\circ$, and $\theta=60^\circ$ away
from the strip normal. Shown are the complete fronts for six
fields $H_a/H_p = 0.2$, 0.4, 0.6, 0.8, 0.93, and 1 (bold line),
where $H_p = (J_c/2)/\sin\theta$, see also Figs.~1-5.
 } \end{figure}   

 \begin{figure}  
\epsfxsize= .98\hsize  \vskip 1.0\baselineskip \centerline{
\epsffile{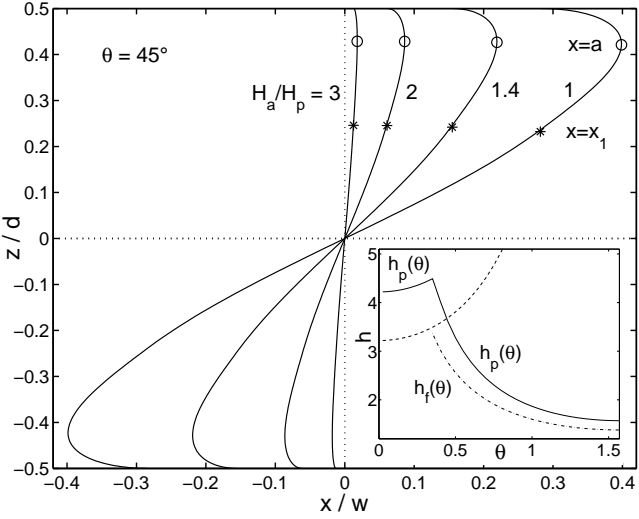}}  \vspace{.1cm}   
\caption{\label{fig7} The current fronts separating the regions
with critical currents of opposite direction in a thin strip in
increasing magnetic field $H_a$ inclined by $\theta=45^\circ$. At
the depicted fields $H_a/H_p = 1$, 1.4, 2, 3 {\it above} the
penetration field $H_p = (J_c/2)/\sin\theta$, the flux-free core
has collapsed into one line, and the front is composed of the
three curves $z_2(0 \le x \le x_1)$ [Eq.~(16)], $z_2(x_1 \le x \le
a)$ [lower branch, Eq.~(15)], and $z_1(a \ge x \ge x_1)$ [upper
branch, Eq.~(11)]. Note that even at large $H_a$ these front lines
{\it are not straight lines} as it might be expected. At $a\ll w$
these fronts collapse into one curve when plotted versus $x/a$.
Inset: the dashed line $h \sim \ln(2ew/d)/\cos\theta $
schematically shows the boundary above which our splitting
procedure fails, see Sec.~II C. The solid line gives the
penetration field $h_p(\theta)\equiv \pi H_p/J_c={\rm
min}[\ln(2ew/d)/ \!\cos\theta ,\, 0.5\pi/\sin\theta ]$ and the
dash-dotted line shows $h_f(\theta)$, Eq.~(\ref{5}).
 } \end{figure}   

 \begin{figure}  
\epsfxsize= .98\hsize  \vskip 1.0\baselineskip \centerline{
\epsffile{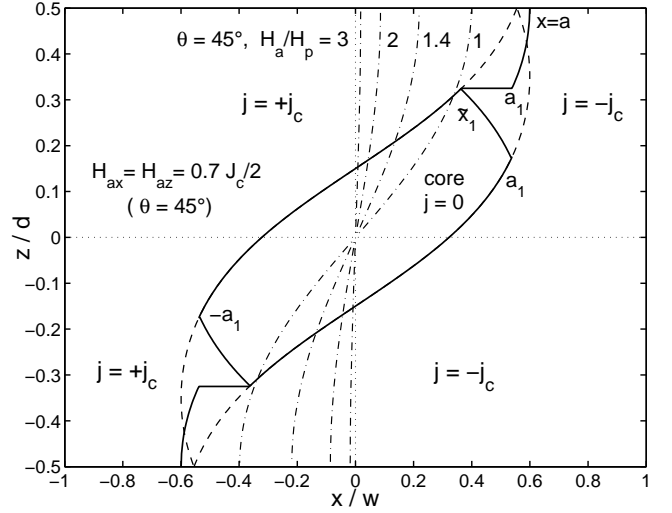}}  \vspace{.1cm}
\caption{\label{fig8} Current and flux fronts in a thin strip to
which first the perpendicular magnetic field $H_{az}$ is applied
and then the in-plane component $H_{ax}$ is increased from zero to
$H_{ax} = H_{az} = 0.7 J_c/2$, resulting in a final tilt angle
$\theta=45^\circ$ (scenario 2, Sec.~III). The solid line
$z_\gamma$ from $x=-a_1$ via $\tilde x_1$, $a_1$, $- \tilde x_1$
to $-a_1$ forms a flux-free core. This core is connected to the
surfaces by tails composed of a horizontal part $z_3(\tilde x_1
\le x \le a_1)$,  and a curved part $z_3(a_1 \le x \le a)$, ending
at the point $(a, d/2)$. The dashed lines give the extensions of
the fronts which exactly coincide with the dashed lines shown in
Fig.~2 and end at $x_0$. The dash-dotted lines are the current
fronts of this scenario above full penetration, at the same fields
as in Fig.~7 for $\theta=45^\circ$. Note that these fronts are
{\it monotonic}, while the corresponding fronts of scenario 1 in
Fig.~7 are S-shaped.
 } \end{figure}   

 \begin{figure}  
\epsfxsize= .98\hsize  \vskip 1.0\baselineskip \centerline{
\epsffile{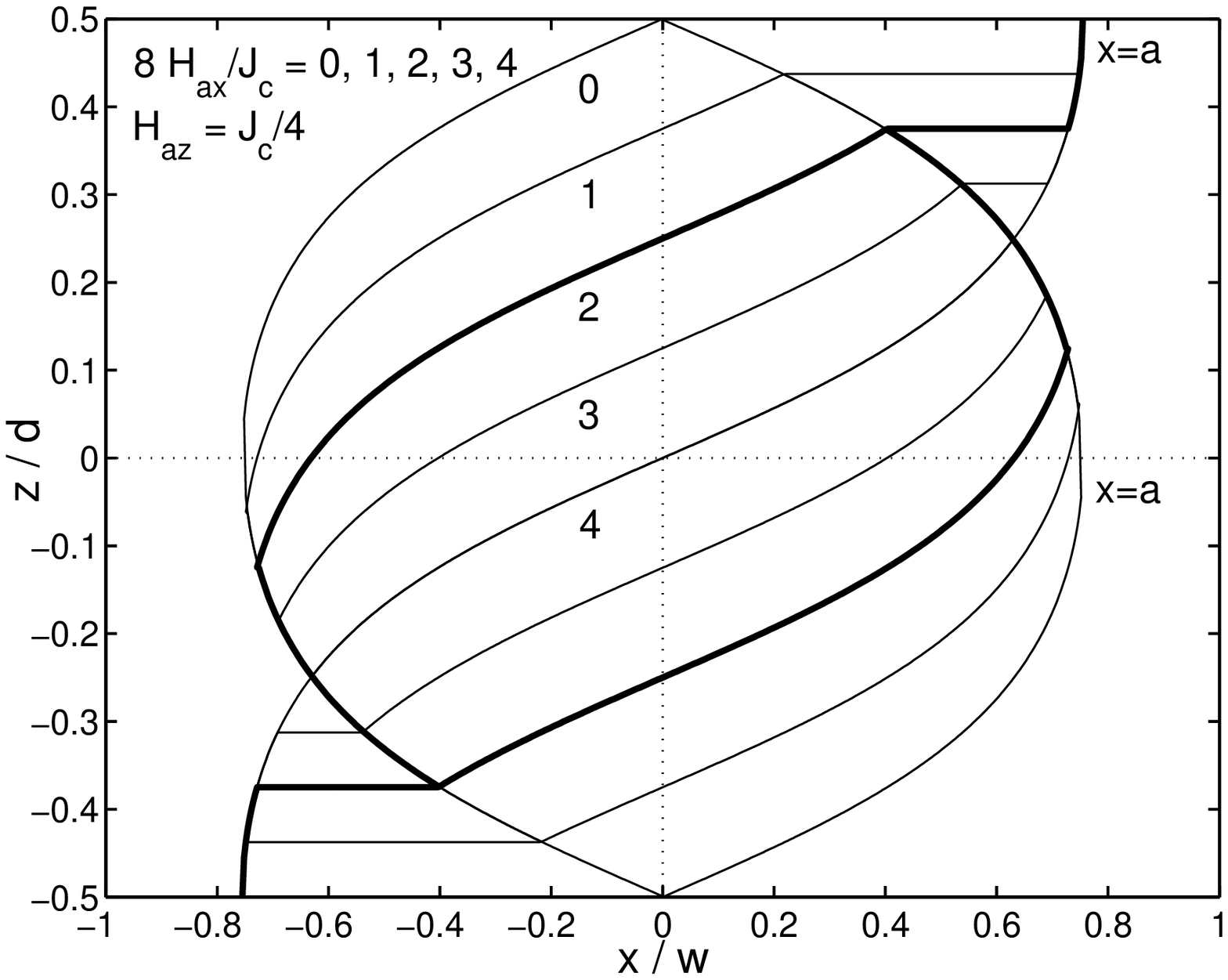}}  \vspace{.1cm}
\caption{\label{fig9} The flux and current fronts in a thin
strip in which first $H_{az} = J_c/4$ is applied and then
$H_{ax}$ is increased from zero to full penetration,
$8 H_{ax} /J_c = 0$, 1, 2, 3, 4 (scenario 2, Sec.~III).
The complete front for $H_{ax} = J_c/4$ is depicted as
bold line, see also Fig.~8. The front 4 through the strip
center $(0,0)$ applies to all $H_{ax} \ge J_c/2$
and depends only on $H_{az}$. Note that the depicted fronts
correspond to different tilt angles $\theta$, e.g., the
saturated front corresponds to $\theta = \arctan(H_{ax}/H_{az})
\ge \arctan 2 \approx 63^\circ$.
 } \end{figure}   

\subsection{Region $h>h_p$}  

Although at $h>h_p$ the $z$ component of the magnetic field does
not penetrate into the region $|x|\le a(h)$ of the strip, its $x$
component completely penetrates into the sample, and the flux-free
core is absent, Fig.~7. The two branches of the boundary
separating the regions of the strip with opposite signs of the
critical current density are still described by Eqs.~(\ref{11})
and (\ref{15}) in the interval $x_1<x<a$. In the interval
$|x|<x_1$ only one of these branches exists, which is the
continuation of the $z_2(x)$ and is described by the formula:
\begin{eqnarray} \label{16}
 z_2(x) &=& {d\over \pi}\ {\rm arctan}
  {x \sqrt{1-a^2} \over \sqrt{a^2-x^2}}
 \nonumber \\  &=& {d\over 4}\ [F(-x,h) -F(x,h) ] \,.
\end{eqnarray}
This formula follows from the given value of the sheet current
$J(x)$, Eq.~(\ref{2}).

Interestingly, even at high magnetic fields $h\gg h_p$ the current
front is S-shaped and does not tend to a straight line as it was
assumed in Ref.~\onlinecite{z}. Probably, it is for this reason
that there is a disagreement between the theoretical and
experimental results in Ref.~\onlinecite{z} at $\theta \sim
\pi/2$. However, it is necessary to keep in mind the following:
Our approximation based on the splitting procedure is valid if
$dz_{\gamma}/dx\ll 1$ and $dz_{1,2}/dx \ll 1$. These inequalities
are fulfilled almost everywhere in the strip when the
characteristic scales in the $x$ direction (i.e., $x_1$, and
$a-x_1$) considerably exceed the thickness $d$. Thus, the region
$h>h_p$ may be considered within our approximation only when
$(\pi/2)\cot\theta < \ln(2ew/d)$, i.e., when the angle $\theta$ is
not too small, see inset in Fig.~7. Otherwise, the component $H_z$
completely penetrates into the sample at lower $H_a$ than the
component $H_x$ does. In this case the field of full penetration
is $H_p^{\perp}/\cos\theta$ where $H_p^{\perp} =
(J_c/\pi)\ln(2ew/d)$ is the penetration field at
$\theta=0$,\cite{17} and the current fronts will differ from those
shown in Fig.~7. Note that the angular dependence of the true
penetration field is given by $H_p(\theta)={\rm
min}(H_p^{\perp}/\!\cos\theta ,\, J_c/2\sin\theta)$ and is a
nonmonotonic function, see inset in Fig.~7.

\section{Magnetic Field Components are Switched on Successively} 

Consider now the scenario when the field $H_{az}$ is switched on
first and then one switches on $H_{ax}$. The resulting magnetic
field again is $H_{ax}=H_a\sin\theta$, $H_{ay}=0$,
$H_{az}=H_a\cos\theta$. In this case the analysis of the critical
state is quite similar to that presented in the Sec.~II, and we
present only the results here.

The field of full penetration of flux is still described by
formula (\ref{4}). However, at any $h\le h_p$ the $x$ size of the
flux-free core, $2a_1$, is less than the width
$2a = 2/\cosh(h\cos\theta)$ of the region where $H_z=0$.
This $a_1$ is determined by the formula $J(a_1)=H_{ax}-J_c$, or
explicitly, by
\begin{equation} \label{17}
 a_1={a\cos( h_x/2) \over
 \left [1-a^2\sin^2( h_x/2)\right ]^{1/2}},
\end{equation}
where $h_x = h \sin\theta = H_{ax}/H_c$. One more characteristic
scale is $\tilde x_1(h)$ determined by the
relation: $J(\tilde x_1)=-H_{ax}$, which leads to the explicit
expression for $\tilde x_1$:
\begin{equation} \label{18}
 \tilde x_1={a\sin( h_x/2) \over
 \left [1-a^2\cos^2( h_x/2)\right ]^{1/2}}.
\end{equation}

In the interval $-a_1\le x\le \tilde x_1$ the shape of the
flux-free core is described by formula (\ref{7}). But in the
region $\tilde x_1\le x \le a_1$ one has
\begin{equation} \label{19}
 z_{\gamma}(x)={d\over 2}\left [1-{2\over\pi}{\rm arctan}
  {x \sqrt{1-a^2} \over \sqrt{a^2-x^2}} \right] .
\end{equation}
In this interval of $x$ there is also a boundary $z_3(x)$
separating the regions of the strip with opposite directions of
$j_c$, see Fig.~8. This horizontal line is described by
\begin{equation} \label{20}
 z_3(x)={d\over 2}\left [1-{h_x \over\pi} \right ] .
\end{equation}
This boundary continues in the region $a_1\le x \le a$ where it is
given by an expression coinciding with Eq.~(\ref{16}):
\begin{equation} \label{21}
 z_3(x)={d\over \pi}\ {\rm arctan}
 {x \sqrt{1-a^2} \over \sqrt{a^2-x^2}} \,.
\end{equation}
When $h$ approaches $h_p$, the point $a_1$ tends to $\tilde x_1$,
and the difference $z_{\gamma}^{\rm upper}(x)-z_{\gamma}^{\rm
lower}(x)$ vanishes simultaneously for all $|x|\le \tilde x_1$. At
$h=h_p$ the flux-free core disappears, while the boundary $z_3(x)$
exists at $h\ge h_p$, and it is described by the expression
(\ref{21}) in the whole interval $-a\le x \le a$, see Figs.~8, 9.
When $H_{ax} \ge J_c/2$ is increased further, this saturated
current front does not change any more.

Thus, we see that the shape of the flux-free core and the boundary
between the regions with opposite directions of the critical
current density do not coincide with those described in Sec.~II
and thus depend on the magnetic history.

 \begin{figure}  
\epsfxsize= .98\hsize  \vskip 1.0\baselineskip \centerline{
\epsffile{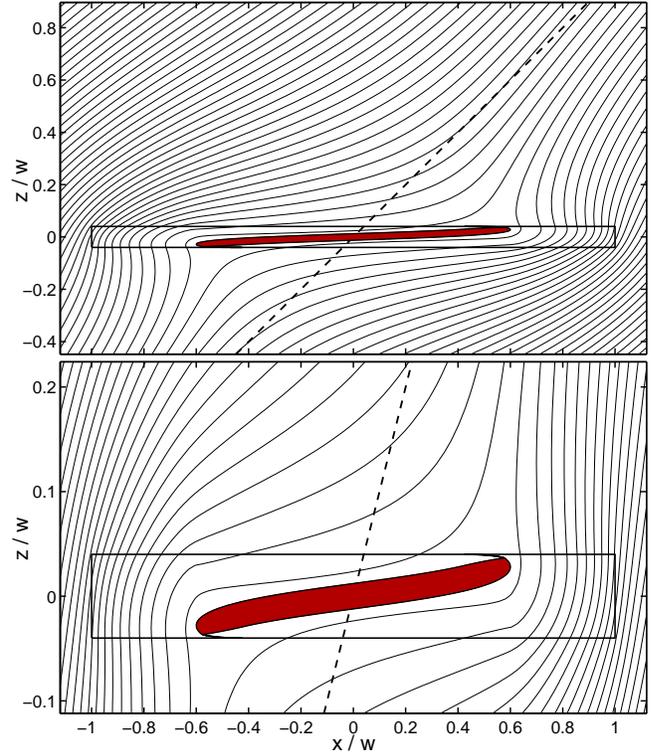}}  \vspace{.1cm}
\caption{\label{fig10} The magnetic field lines for a strip of
width $2w$ and thickness $d=0.08 w$ in an applied field inclined
by $\theta = 45^\circ$ and increased from zero to $H_a = 0.7 H_p$
as in Fig.~2 (scenario 1). These field lines were computed as
equidistant contour lines of the vector potential $A_y(x,z)$,
Eq.~(A1), yielding a density of lines proportional to the local
magnetic field $H(x,z)$. The gray area shows the current and
field-free core. The dashed line is along $H_a$. The upper plot is
to scale, the lower plot is stretched along $z$ by a factor of 4.
 } \end{figure}   

 \begin{figure}  
\epsfxsize= .98\hsize  \vskip 1.0\baselineskip \centerline{
\epsffile{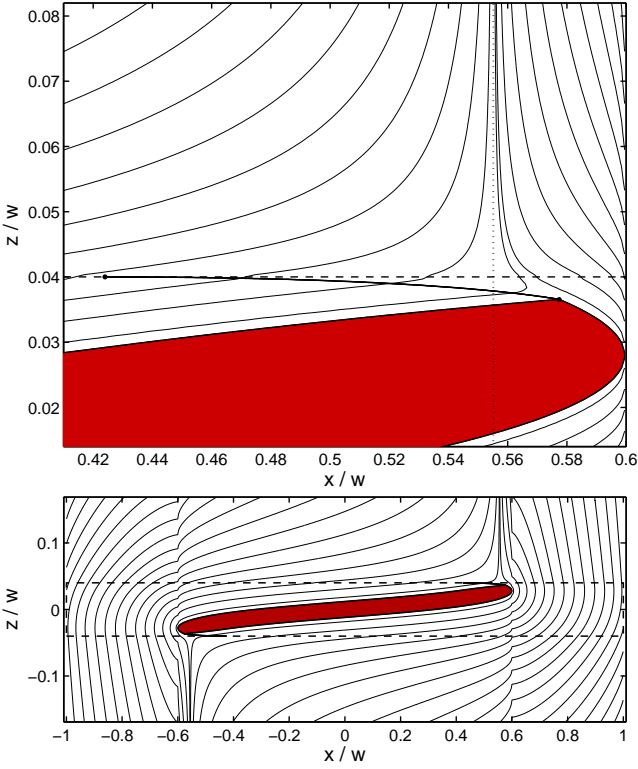}}  \vspace{.1cm}  
\caption{\label{fig11} The magnetic field lines for the same strip
as in Fig.~10 ($d=0.08 w$, $\theta = 45^\circ$, $H_a = 0.7 H_p$)
from Eqs.~(\ref{A2})-(\ref{A7}). To reveal the details near the
current fronts, the field lines here are nonequidistant contour
lines of the analytic vector potential $A_y(x,y)$ of Appendix A,
with levels $A_\nu \propto \nu^2 {\rm sign}(\nu)$, $\nu = 0$, $\pm
1$, $\pm 2, \dots$, yielding more lines at low fields. The gray
area shows the current and field-free core. The dashed line marks
the surface of the strip and the dotted line shows $x=x_0$,
Eq.~(\ref{8}), defined by $H_x(x_0, d/2)=0$. Note that field lines
cut the line (from $x=x_1$ to $x=x_2$ marked by bold dots) that
separates regions with $j_y=\pm j_c$. The lower plot shows the
same case on different $x$ and $z$ scales.
 } \end{figure}   

 \begin{figure}  
\epsfxsize= .98\hsize  \vskip 1.0\baselineskip \centerline{
\epsffile{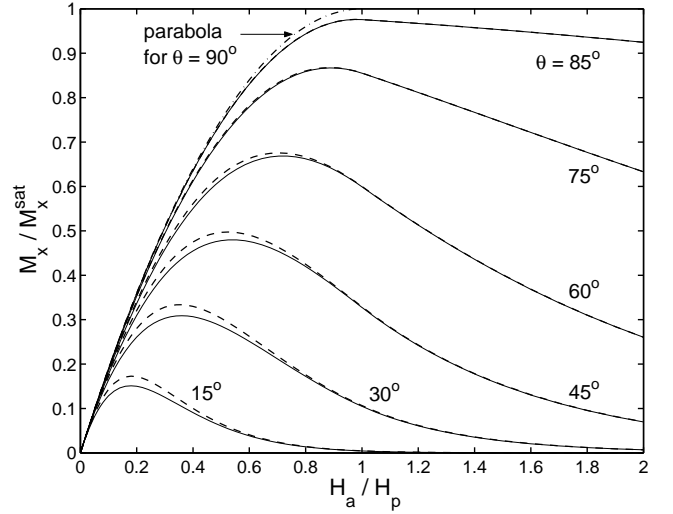}}  \vspace{.1cm}
\caption{\label{fig12} Magnetic-moment component $M_x(H_a) /
M_x^{\rm sat}$ from Eq.~(\ref{24}) and Appendix B, for scenario 1
(solid lines) and scenario 2 (dashed lines) plotted versus the
applied magnetic field $H_a$ in units of the penetration field
$H_p = J_c/2\sin\theta$ for tilt angles $\theta = 15^\circ$,
$30^\circ$, $45^\circ$, $60^\circ$, $75^\circ$, and $85^\circ$.
Here $M_x^{\rm sat} = -J_cdw/2$. The dot-dashed parabola
$M_x/M_x^{\rm sat} = 1-(1-H_a/H_p)^2$ applies to $\theta=90^\circ$
($H_a$ along $x$), and for $\theta=0$ ($H_a$ along $z$) one has
$M_x=0$. For all other angles the $|M_x(H_a)|$ for scenario 2 is
larger than for scenario 1.
 } \end{figure}   

 \begin{figure}  
\epsfxsize= .98\hsize  \vskip 1.0\baselineskip \centerline{
\epsffile{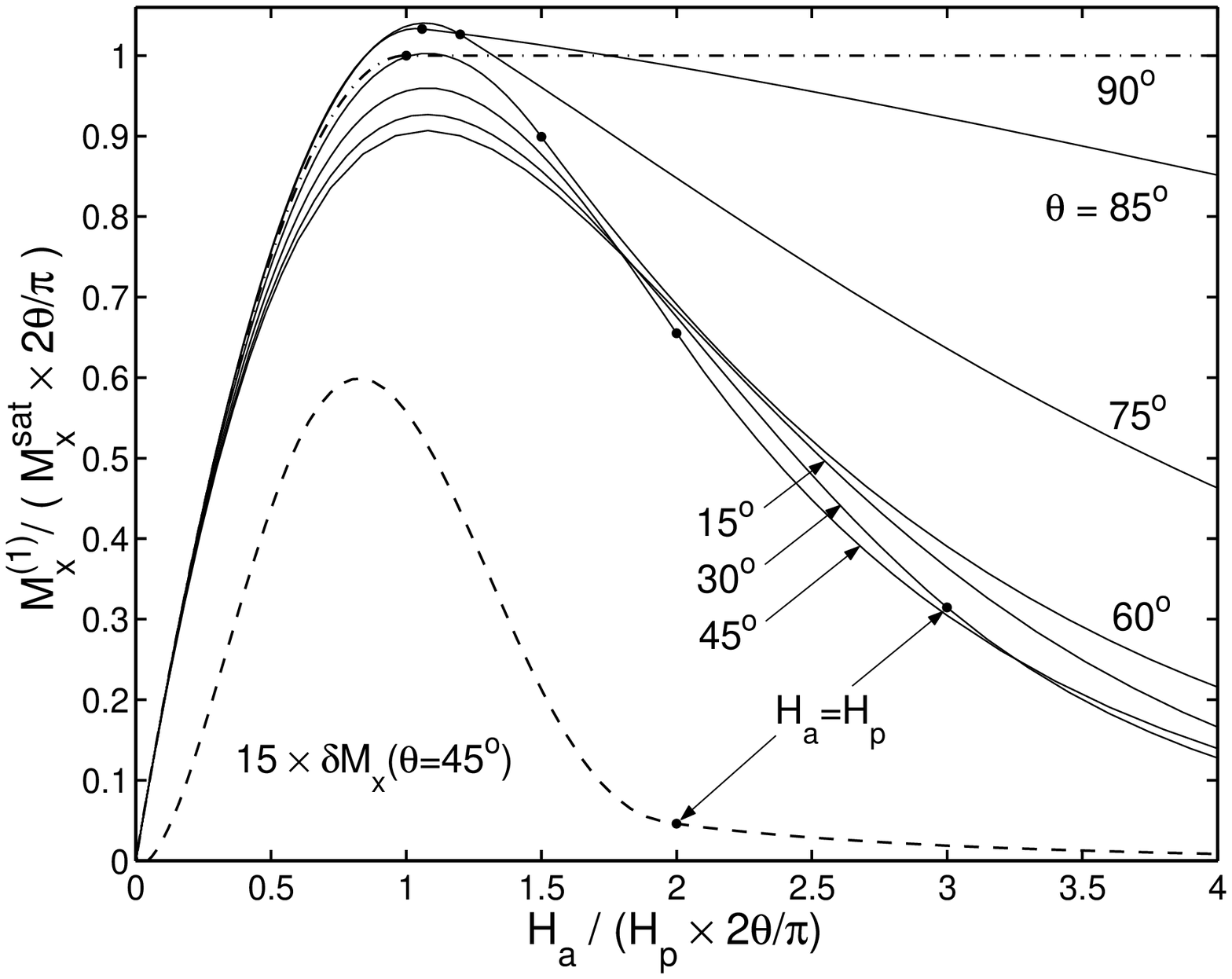}}  \vspace{.1cm}
\caption{\label{fig13} The same magnetization curves as in Fig.~12
for scenario 1, plotted as $M_x/ (M_x^{\rm sat}\times
2\theta/\pi)$ versus $H_a/ (H_p \times 2 \theta / \pi)$. The dots
mark the point where $H_a = H_p$. The dot-dashed curve gives the
limit $\theta=90^\circ$. The dashed line shows the difference of
the $M_x$ of scenarios 2 and 1 for $\theta = 45^\circ$ in form of
$15\times (M_x^{(2)} -M_x^{(1)}) /(M_x^{\rm sat} \times
2\theta/\pi)$.
 } \end{figure}   

\section{Magnetic field lines}

Using the obtained results, it is easy to find the distribution of
the magnetic fields in the critical state of the strip. One may
either integrate over the current-carrying area, noting that each
current path has the magnetic field of a straight wire (Fig.~10).
Or one may derive analytical expressions using our splitting
approximation (Fig.~11). The $z$ component, $H_z(x)$, in this
approximation does not depend on $z$ and is given by the formulas
of Refs.~\onlinecite{2a,3,4} [or by Eq.~(A8) in compact form], and
the $x$ component is
\begin{equation}\label{22}
 H_x(x,z)=H_x(x,-d/2)+\int_{-d/2}^z\! j_y(x,z') dz' ,
\end{equation}
where $H_x(x,-d/2)$ is the field on the lower surface of the
strip. At $|x|\le a$ one has $H_x(x,-d/2)=(J_c/2)F(-x,h)$, while
$H_x(x,-d/2)=H_a\sin\theta+(J_c/2){\rm sign}(x)$ at $a\le |x|\le
1$. Taking into account that $j_y(x,z)$ has only the values $j_c$,
$-j_c$, or $0$, and knowing $z_{\gamma}(x)$ and the boundaries
between the regions with $\pm j_c$, one can easily calculate
$H_x(x,z)$ everywhere in the strip and near the strip explicitly.

As an example, Figs.~10 and 11 show the magnetic field lines
(parallel to the Abrikosov vortex lines) in the strip and near the
strip for the first scenario of switching on the magnetic field at
constant tilt angle $\theta$ till $H_a= 0.7 H_p$ is reached. Both
figures show the field lines obtained as contour lines of the
vector potential $A_y(x,z)$ related to ${\bf H}(x,z) = \nabla
\times ({\bf \hat y} A_y)$. Figure~10 depicts the field lines
calculated directly from Amp\` ere's law using the currents
obtained in Sec.~II A and formula (\ref{A1}) of Appendix A at $d/w
= 0.08$. It is important that this current distribution indeed
leads to a flux-free core which is close to that obtained in
Sec.~II A. On the other hand, figure 11 uses expressions
(\ref{A2}) - (\ref{A8}) of Appendix~A for the same $d/w=0.08$.
These expressions were derived with our splitting procedure. It
can be seen that the agreement of both field-line patterns is
good, but the fine details near the current fronts can be more
easily resolved in Fig.~11 (top) which is based on the simple
analytical formulas. In particular one can see that the field
lines exactly flow around the core in which $j=0$, and some field
lines cut the line (``tail'') that separates regions with $j_y=
\pm j_c$ and runs from $x=x_1$ on the upper surface to the cusp of
the core at $x=x_2$, see also Fig.~2. The slight wiggle of the
field lines occurring near $|x|=a$ in Fig.~11 (bottom) is an
artifact, since the condition for the splitting procedure,
$dz_{\gamma}/dx \ll 1$, fails at this point. However, a more
detailed analysis shows that the difference between the field-line
patterns of Figs.~10 and 11 manifests itself only in narrow
intervals near $x = \pm a$.

\section{Magnetic moment}

In an oblique magnetic field, apart from the $z$ component of the
magnetic moment of the strip, $M_z$, an $x$ component $M_x$
appears, and both can be investigated experimentally.\cite{z}
The expression for $M_z$ (per unit length along $y$) is
known:\cite{3}
\begin{equation} \label{23}
 m_z\equiv -{M_z\over J_cw^2}=\tanh (h\cos\theta)\,.
\end{equation}
Knowing $z_{\gamma}(x)$ and the boundaries between the regions
with opposite signs of the critical current density, given in
Secs.~II and III, one can calculate $M_x$ for the strip from the
formula:
\begin{equation} \label{24}
 M_x=-\int_{-a}^a\! dx\int_{-d/2}^{d/2}\!\! zj_y(x,z) dz \,.
\end{equation}
Here $M_x$ is given per unit length along $y$, and we have taken
into account that only the region of the strip, $|x|\le a$, gives
a nonzero contribution to $M_x$. The saturation value of $M_x$,
which is achieved in high fields for $\theta=\pi/2$, is $M_x^{\rm
sat} = -J_cdw/2$, see Appendix B.

It is known for the infinitely thin strip \cite{5} that whatever
its length in $y$, the ends of the strip (in $y$) always give the
same contribution to $M_z$ as that caused by the currents $j_y$
flowing in the $y$ direction. When $M_x$ is calculated, it is
necessary to allow for the fact that near the ends of the strip
the currents may have not only $y$ and $z$ components but also an
$x$ component, i.e., the problem becomes three dimensional.
However, using the conservation law for the current, ${\rm
div}{\bf j}=0$, one can show that even in this three dimensional
case the ends of the strip strictly double $M_x$. It is for this
reason that the factor $1/2$ was omitted in formula (\ref{24}).

  In Fig.~12 we compare the $H_a$-dependences of $M_x$ for the two
scenarios of switching on the magnetic field. Note that
$M_x^{(2)}(H_a)$ of scenario 2 is always larger than
$M_x^{(1)}(H_a)$ of scenario 1, except for the trivial angles
$\theta =0$ where $M_x=0$, and $\theta = \pi/2$ where $M_x /
M_x^{\rm sat} = 1 - (1-H_a/H_p)^2$. All $M_x(H_a)$ exhibit a
maximum of height $M_x^{\rm max}/M_x^{\rm sat} \approx
2\theta/\pi$ occurring at $H_a / H_p \approx 2 \theta/\pi$, and
they have the same slope $\partial M_x / \partial H_a = 2M_x^{\rm
sat}/H_p$ at $H_a=0$. The difference $M_x^{(2)}-M_x^{(1)}$ is also
maximum near $H_p$, see the dashed curve in Fig.~13. Figure 13
plots the $M_x^{(1)}$ of scenario 1 as in Fig.~12 but with both
abscissa and ordinate stretched by a factor $\pi/(2\theta) \ge 1$
such that the approximate scaling of the $M_x^{(1)}(H_a)$ at not
too large $H_a/H_p$ is seen. The $M_x^{(2)}(H_a)$ curves of
scenario 2 scale even better.

The nonmonotonic dependence of $M_x$ on $H_a$ can be
understood from the following arguments: In the region
$|x|<a$ the $x$ component of the external magnetic field,
$H_{ax}=H_a\sin\theta$, leads to an asymmetric distribution of
the currents over $z$, see Figs.~1-9. It is this asymmetry that
generates the $M_x$ component. Thus, $M_x$ can be estimated as
follows:
$M_x\sim M_x^{\rm sat}\cdot (a/w)(H_a/H_p)$. The factor $a/w$
decreases with $H_a$, see Eq.~(\ref{1}), and its product with
the increasing factor $H_a/H_p$ leads to the observed
nonmonotonic behavior of $M_x(H_a)$.

\section{Conclusions}

We solve the critical state problem for a strip of finite
thickness in an oblique magnetic field. Two scenarios of switching
on the external magnetic field are considered: (1) the magnetic
field is increased at a constant tilt angle $\theta$, and (2) the
magnetic field components are switched on successively. The
resulting critical states are different in these two cases,
{\it even after the flux has fully penetrated the strip}.

  Another characteristic feature of both states is that, below
the field of full penetration, the height of the flux-free core is
{\it less} than the strip thickness, i.e., the core does not reach
the flat surfaces but is connected to them by lines (``tails'')
that separate regions with opposite direction of the critical
currents.\cite{16}  Moreover, the width of the core may be
narrower than the region of the strip in which the $z$ component
of the magnetic field (i.e., the component perpendicular to the
plane of the strip) vanishes.

  One more interesting feature of the
critical states in strips in an oblique magnetic field follows
from the data of Figs.~6 and 9: When the applied field increases,
and hence the flux lines further penetrate into the sample, the
current distribution changes not only near the flux front but also
away from it, i.e., in the regions where the critical state was
established before. Note that this feature is also seen in
figures of Ref.~\onlinecite{Camp} in which the critical state of a
rotating cylinder was considered in magnetic fields
perpendicular to its axis. These findings mean that this property
of the current distributions in the critical states is
characteristic of the general case when the geometry of experiment
is not too symmetric, while the usual change of currents at the
penetrating flux front occurs only in special symmetric situations
(e.g., when the tilt angle $\theta=0$, $\pi/2$, or when the
cylinder does not rotate). Finally, we found the somewhat
unexpected result that in an oblique magnetic field $H_a$
exceeding the penetration field, the current front separating the
regions of the strip with $\pm j_c$ generally is {\it not a straight
line} and can shift in the sample when $H_a$ is increased further.

The above general features of the critical sate are robust and
hold even if the strip is not very thin or if its cross section
is not rectangular, i.e., these features are common for all
critical states in {\it nonsymmetric} situations. However, the fine
details of our results (e.g., the short tail from $x_1$ to $x_2$
in Fig.~2) require that all characteristic lengths in the plane
of the strip considerably exceed its thickness. If some length of
the flux-free core does not satisfy this condition, one may expect
deviations from the presented results in this region of the core.
Furthermore, the temperature should be low enough that flux creep
does not smear these details, i.e., the creep exponent $n$ in
the current--voltage law $E(j) \propto (j/j_c)^n$ should be
large. A detailed numerical investigation of the effect of flux
creep is under way. Note that the detailed shape of the flux-free
core and of the boundaries between regions with opposite critical
current in principle can be investigated via the in-plane
component of the magnetic moment, while it has little influence
on the magnetic field on the surface.

\acknowledgments

  This work was supported by the German Israeli Research
Grant Agreement (GIF) No G-705-50.14/01 and by the
European INTAS project 01-2282.

\appendix
\section{Vector potential}  

  The magnetic field lines of a strip parallel to $y$ coincide
with the contour lines of the vector potential $A_y(x,z)$ related
to the current density $j_y(x,z)$ by $\nabla^2 A_y = -j_y$ or
  \begin{equation}\label{A1}
  A_y({\bf r}) = \int \!\! d^2r'\, j_y({\bf r}') \,
             { \ln |{\bf r - r}'| \over 2\pi} \,,
  \end{equation}
with ${\bf r} = (x,z)$.

For the special case of a thin strip in an oblique applied field,
we can also find $A_y(x,z)$ using our splitting procedure. Here,
as an example, we give the expressions for $A_y$ in the case of
scenario 1 at $h \le h_f$ (see Fig.~2). Inside the core delimited
by the two lines $z_\gamma(x)$ and $-z_\gamma(-x)$, one has
$H_x=H_z=j_y=0$, and we may put $A_y=0$ there. Within the core
width ($|x|<a$) one obtains:
  \begin{equation}\label{A2}
  A_y(x,z) = -\int_{z_\gamma(x)}^z \!\! H_x(x,z')\, dz'.
  \end{equation}
Inserting the $H_x$ from Sec.~IV into Eq.~(\ref{A2}), one finds
explicit formulas for $A_y$. Equivalently, the $A_y$ inside
the core width can be calculated directly from the equation
$\partial^2 \! A_y /\partial z^2 = - j_y$ and the current density
of Sec.~II. Eventually, we obtain the following expressions for
$A_y = (j_c/2)\, a_y(x,z)$, depending on $a$, $x_1$, $x_2$,
$z_\gamma(x)$, and $z_1(x)$: For $-a < x < x_1$, $z_\gamma < z <
d/2$ and for $x_1 < x < x_2$, $z_\gamma < z < z_1$:
  \begin{equation}\label{A3}
  a_y = 2A_y /j_c = -(z - z_\gamma)^2  \,,
  \end{equation}
for $x_1 < x < x_2$, $z_1 < z < d/2$:
 \begin{equation}\label{A4}
  a_y = (z\!-\!z_1)(z\!-\!3z_1\!+2z_\gamma)
     -\!(z_1\! - \!z_\gamma)\!^2 \,,
  \end{equation}
for $x_2 < x < a$, $z_\gamma < z < d/2$:
  \begin{equation}\label{A5}
  a_y =  (z - z_\gamma)^2  \,.
  \end{equation}
For $-a < x < a$, $z\ge d/2$ (above the strip) one has
  \begin{equation}\label{A6}
  A_y(x,z) = A_y(x,d/2) -(J_c/2) F(x,h) (z -d/2).
  \end{equation}
The appropriate expressions below the core follows from the
symmetry relationship $A_y(x,-z) = -A_y(-x,z)$. Outside the core
width one has for $x>a$ and all $z$ inside or close to the strip
[for $x <-a$ use $A_y(x,z)=-A_y(-x,-z)$]:
  \begin{eqnarray}\label{A7}
  A_y(x,z) = A_y(a,0) + \! \int_a^x\!\!\! H_z(x',0) dx'
    \! -\!H_{ax}z   \nonumber \\  +  {J_c\over 2} g(x,z)
  \end{eqnarray}
where $A_y(a,0)=d\,H_{ax}^2/2J_c$, $g=z^2/d$ inside and $g= |z|\!
-\! d/4$ outside the strip for $x\! <\! w$, $g = 0$ for $x\!
>\! w$, and $H_z(x,0)$ from Ref.~\onlinecite{2a}. Using formulas ${\rm
arctanh}(1/u) = {\rm arctanh}(u) + i\pi/2$ (at $|u| < 1$) and
${\rm arctanh}(iu) =\! i\,{\rm arctan}(u)$, we may write one
single expression valid for all $-\infty <x <\infty$ (Re means
real part):
  \begin{equation}\label{A8}
  H_z(x,0) = {J_c\over \pi} {\rm Re}\bigg\{ {\rm arctanh}
  \sqrt{ 1-a^2/x^2 \over 1 -a^2}\, \bigg\} .
  \end{equation}

\section{Magnetic moment}   

  From the general definition of the magnetic moment of the strip
per unit length, $(M_x, M_z) = \int dx \int dz (-z, x)J_y(x,z)$,
one obtains the following saturation values in the two limiting
cases: For $H_a \ge H_p^\perp = (J_c/\pi) \ln(2ew/d)$ \cite{17}
along $z$, one has $j(x,z) = -j_c\, {\rm sign}(x)$ and $M_z =
M_z^{\rm sat}= -j_c d w^2 = -J_c w^2$. For $H_a \ge J_c/2$ along
$x$, one has $j(x,z) = j_c\, {\rm sign}(z)$ and $M_x = M_x^{\rm
sat}= -j_c d^2 w/2 = -J_c d w/2$. The reduced magnetic moment $m_z
= M_z /M_z^{\rm sat}$ along $z$ depends only on
$H_{az}=H_a\cos\theta$, and is given by Eq.~(\ref{23}). The
magnetic moment along $x$, Eq.~(\ref{24}), in general depends on
both components and has to be computed from the current fronts.
Explicit formulas for $m_x = M_x /M_x^{\rm sat}$ can be obtain for
any $H_a$, but here we present them only in the case $H_a \ge H_p
= J_c/2 \sin\theta$. Namely, for scenario 1 the formulas of
Sec.~II yield for $h \ge h_p$
  \begin{eqnarray}\label{B2}
  m_x = \int_0^{x_1}\!\!\! dx \left[ 1- \bigg({2\over \pi}
  {\rm arctan} {x\sqrt{1-a^2} \over
  \sqrt{a^2-x^2}} \bigg)^{\!\!2} \right] \nonumber \\
  + \int_{x_1}^a\!\! dx \left[ 1- {2\over \pi}
  {\rm arctan} {x\sqrt{1-a^2} \over
  \sqrt{a^2-x^2}} \,\right]  \nonumber \!\times \\
  \left[ 1\!+ \!{2\over \pi}(h\!-\!h_*)\sin\theta +\!{2\over \pi}
  {\rm arctan} {\sqrt{\sin^2\!\theta\! -\! x^2} \over
   \cos\theta } \,\right] ,
  \end{eqnarray}
with $h_* = {\rm arcosh}(\sin\theta/x) / \cos\theta$, Eq.~(9).
For scenario 2 the formulas of Sec.~III yield for $h \ge h_p$
  \begin{eqnarray}\label{B3}
  m_x = \int_0^{a}\!\! dx \left[ 1- \bigg({2\over \pi}
  {\rm arctan} {x\sqrt{1-a^2} \over
  \sqrt{a^2-x^2}} \bigg)^{\!\!2} \right] ,
  \end{eqnarray}
which does not depend on $H_{ax}$.

{}


\begin{thebibliography}{}
\bibitem{16} A detachment of the core from the surface,
and the appearance of two straight lines separating regions with
$j= \pm j_c$, occurs even in perpendicular field ($\theta=0$) when
the applied field is so large that the core width $2a$ becomes
smaller than the strip thickness $d$, namely, at $H_a\approx 0.5
H_p^\perp - H_p^\perp$ where $H_p^\perp = (J_c/\pi)\ln(2ew/d)$ is
the field of full penetration of a perpendicular field \cite{17},
see Fig.~2 in Ref.~\onlinecite{17}. In contrast, our present
result for $\theta \gg d/w$ yields core detachment at arbitrarily
large core width.
\end{thebibliography}
\end{document}